\documentclass[aps,prl,reprint,superscriptaddress]{revtex4-2}
\usepackage[T1]{fontenc}
\usepackage[utf8]{inputenc}
\usepackage{amsmath,amssymb,amsthm,bm,mathtools}
\usepackage{graphicx}
\usepackage{hyperref}
\usepackage{microtype}
\usepackage{braket}
\usepackage{float}
\usepackage{xcolor}
\hypersetup{colorlinks=true,citecolor=blue,linkcolor=blue,urlcolor=blue}

\newcommand{\amend}[1]{\textcolor{black}{#1}}
\newcommand{\tr}{\operatorname{Tr}}
\newcommand{\ii}{\mathrm{i}}
\newcommand{\dk}{\hat{\mathbf d}}
\newcommand{\bsigma}{\boldsymbol\sigma}
\newcommand{\hc}{\mathrm{H.c.}}

\begin{document}

\title{Exact Criteri\amend{a} for Ground-State Overlap Dominance \amend{and Dynamical Quantum Phase Transitions}}

\author{Taisanul Haque}
\affiliation{%
	Institute for Theoretical Physics, \href{https://ror.org/01y9bpm73}{University of G\"ottingen}, Friedrich-Hund-Platz 1, 37077 G\"ottingen, Germany
	}	
\email{taisanul.haque@stud.uni-goettingen.de}

\date{\today}

\begin{abstract}
	It was recently conjectured and verified for the transverse-field Ising model [Phys. Rev. B 113, 165102 (2026)] that, after a sudden quench within the same equilibrium phase, the initial ground state has its largest overlap with the final ground state. We show that this phase-based criterion is generally false, even in translationally invariant free-fermion systems. For Hamiltonians that factorize into independent $2\times 2$ momentum sectors, we derive the exact necessary-and-sufficient condition for ground-state overlap dominance: the initial and final sector Bloch vectors must have positive dot product for every momentum. This result proves the conjecture in classes where same-phase quenches enforce this geometric condition, but gives explicit same-phase counterexamples in Kitaev chains, where excited final eigenstates can dominate the overlap distribution. We further show that the same obstruction controls real-time Fisher-zero crossings, allowing dynamical quantum phase transitions without crossing an equilibrium phase boundary.
\end{abstract}

\maketitle

The adiabatic theorem gives the reference point for spectral ordering under slow driving. In quantum mechanics, this line begins with the work of Born and Fock and was later put into a modern operator form by Kato. It has since been extended beyond the standard gap assumption in several directions \cite{BornFock1928,Kato1950,AvronElgart1999}. A sudden quench probes the opposite limit. One prepares the ground state of an initial Hamiltonian, changes a control parameter on a short time scale, and then asks how that state decomposes in the eigenbasis of the final Hamiltonian. This problem is central in nonequilibrium many-body physics because it fixes the spectral weights before one studies local observables, correlation functions, relaxation, or late-time ensembles \cite{CalabreseCardy2006,Cazalilla2006,RigolDunjkoOlshanii2007,Polkovnikov2011,Dziarmaga2010,EsslerFagotti2016}.

Quantum quenches matter both theoretically and experimentally. They organize exact results in integrable chains, motivate generalized Gibbs descriptions, and underlie light-cone spreading, entanglement growth, and thermalization in isolated systems \cite{RigolDunjkoOlshanii2007,Polkovnikov2011,EsslerFagotti2016,Cheneau2012,Jurcevic2014,Kaufman2016}. For the present problem, the central object is the set of overlap weights
$ p_A=|\langle E_A^{(f)}|\Psi_0^{(i)}\rangle|^2$, 
where $|\Psi_0^{(i)}\rangle$ is the initial ground state and $|E_A^{(f)}\rangle$ are final many-body eigenstates. These weights enter both the work distribution\cite{TalknerLutzHanggi2007, Silva2008,Goold2019, Rylands2019} and, more directly for the present purpose, the Loschmidt amplitude
\begin{equation}
 \mathcal G(t)=\langle \Psi_0^{(i)}|e^{-\ii H_f t}|\Psi_0^{(i)}\rangle=
 \sum_A p_A e^{-\ii E_A^{(f)} t}
\end{equation}
so overlap ordering is naturally tied to Loschmidt physics and dynamical quantum phase transitions(DQPTs)  \cite{HeylPolkovnikovKehrein2013,Heyl2018}. \amend{Complementary studies have characterized overlap distributions and related spectral content generated by quenches in interacting lattice models and integrable field theories \cite{Mazza2016, Rylands2019}, while general field-theoretic analyses have related quasiparticle or energy-eigenstate expansions of nonequilibrium states to their subsequent unitary dynamics \cite{DelfinoSorba2022,DelfinoSorba2024}}. It also connects naturally to the equilibrium systems on fidelity and ground-state overlaps as probes of phase structure \cite{ZanardiPaunkovic2006,Gu2010,CozziniZanardiPazzi2007}.

Exact free-fermion chains are the natural setting in which this question can be solved without uncontrolled approximations. The Ising chain becomes quadratic after the Jordan-Wigner and Bogoliubov transformations \cite{LiebSchultzMattis1961,Pfeuty1970,CalabreseEsslerFagotti2011,CalabreseEsslerFagotti2012}. Standard two-band and Bogoliubov-de Gennes models such as the \amend{Su--Schrieffer--Heeger (SSH) chain}, and the Kitaev chain have the same sectorwise structure \cite{SSH1979,Kitaev2001}. A recent work by Damerow and Kehrein asked whether quenches that remain inside one phase should always have their largest overlap with the final ground state, proved this for the transverse-field Ising chain \amend{(TFIM)}, and gave partial evidence for the \amend{axial next-nearest-neighbor Ising }(ANNNI) chain \cite{DamerowKehrein2025}. The open issue is whether the mechanism is truly phase based, or whether the Ising result relies on a stronger hidden geometric constraint.

In this Letter we solve the ordering problem exactly for a broad class of Hamiltonians. For translationally invariant free-fermion models that factorize into independent $2\times2$ sectors, the many-body problem reduces to a sectorwise condition on Bloch vectors. The condition is necessary and sufficient. It yields a simple affine-family corollary and makes the theorem-valid region explicit. Throughout, ``same phase'' means the same connected gapped region on the chosen one-parameter line. For the TFIM and the SSH chain, the theorem-valid region coincides with the full physical same-phase region. For the Kitaev chain, this coincidence does not hold in general. We therefore obtain explicit quenches that remain within a single physical phase but nevertheless violate the phase-only rule. \amend{The same sectorwise positivity condition also gives an exact no-go result for DQPTs: when the initial and final sector Bloch vectors have a positive dot product at every momentum, Fisher-zero lines cannot cross the real-time axis. The converse does not follow. Violating the positivity condition only makes a DQPT possible; an actual crossing requires the initial and final sector Bloch vectors to become orthogonal in at least one momentum sector. Our finite-range example realizes this orthogonality condition within a single connected equilibrium phase.
}

\paragraph*{Model and exact theorem.}
We consider translationally invariant free-fermion Hamiltonians of the form
\begin{equation}
H(\lambda)=\sum_{k\in\mathcal K}\chi_k^\dagger
\Bigl[\epsilon_0(k,\lambda)\,\mathbb{I}+\mathbf d(k,\lambda)\cdot\bsigma\Bigr]\chi_k
\label{eq:H}
\end{equation}
where $\chi_k$ is a two-component fermionic spinor, $\mathbf d(k,\lambda)\in\mathbb R^3$, and $k$ labels an independent crystal-momentum sector or an independent $(k,-k)$ Bogoliubov sector. We assume the region of interest is gapped, $|\mathbf d(k,\lambda)|>0 ,\; \forall\,k,\lambda.$
For a quench $\lambda_i\to\lambda_f$, define the normalized Bloch vectors
\begin{equation}
\dk_{\alpha,k}=\frac{\mathbf d(k,\lambda_\alpha)}{|\mathbf d(k,\lambda_\alpha)|},
\qquad \alpha=i,f
\end{equation}
and the sectorwise dot product $x_k:=\dk_{i,k}\cdot\dk_{f,k}.$
The occupied and empty projectors in sector $k$ are
\begin{equation}
P^-_{\alpha,k}=\frac{1-\dk_{\alpha,k}\cdot\bsigma}{2},
\qquad
P^+_{\alpha,k}=\frac{1+\dk_{\alpha,k}\cdot\bsigma}{2}
\end{equation}
The initial many-body ground-state projector is $P_i=\bigotimes_k P^-_{i,k}$. A final many-body eigenstate is labeled by a subset $A\subset\mathcal K$: sectors in $A$ are excited to the upper level, and sectors outside $A$ stay in the lower level. Its projector is
\begin{equation}
P_A^{(f)}=\bigotimes_{k\notin A}P^-_{f,k}\bigotimes_{k\in A}P^+_{f,k}
\end{equation}
The overlap weight is $p_A:=\tr(P_i P_A^{(f)})$. Using $(\hat{\mathbf a}\cdot\bsigma)(\hat{\mathbf b}\cdot\bsigma)
=(\hat{\mathbf a}\cdot\hat{\mathbf b})\mathbb{I}
+\ii(\hat{\mathbf a}\times\hat{\mathbf b})\cdot\bsigma$
and $\tr\bsigma=0$, one obtains
\begin{equation}
\tr(P^-_{i,k}P^-_{f,k})=\frac{1+x_k}{2},
\qquad
\tr(P^-_{i,k}P^+_{f,k})=\frac{1-x_k}{2}
\end{equation}
Because the sectors are independent,
\begin{equation}
 p_A=
\prod_{k\notin A}\frac{1+x_k}{2}
\prod_{k\in A}\frac{1-x_k}{2}
\label{eq:master}
\end{equation}
The final ground state corresponds to $A=\varnothing$, with weight
$ p_0=\prod_k\frac{1+x_k}{2}$. For any nonempty $A$,
\begin{equation}
\frac{p_A}{p_0}=\prod_{k\in A}\frac{1-x_k}{1+x_k}
\label{eq:ratio}
\end{equation}
Hence the ordering criterion is immediate.

\emph{Theorem.} For the factorized free-fermion Hamiltonian \eqref{eq:H}, the final ground state is the unique maximal-overlap final eigenstate after the quench $\lambda_i\to\lambda_f$ if and only if
\begin{equation}
\dk_{i,k}\cdot\dk_{f,k}>0\qquad \forall\,k\in\mathcal K
\label{eq:criterion}
\end{equation}
\amend{Here, maximal overlap means the largest individual many-body overlap weight $(p_A)$ among the final eigenstates. Now,} if the inequalities are only nonstrict, the final ground state is still a maximizer, but it need not be unique.

The proof is direct from \eqref{eq:ratio}. If $x_k>0$ for every $k$, then every factor on the right-hand side lies in $[0,1)$, so $p_A<p_0$ for all nonempty $A$. If one sector has $x_{k_0}<0$, then the one-sector excitation $A=\{k_0\}$ satisfies $p_A>p_0$. If $x_{k_0}=0$, then $p_{\{k_0\}}=p_0$.

A useful corollary holds for affine one-parameter families,
\begin{equation}
\mathbf d(k,\lambda)=\mathbf a_k+\lambda\,\mathbf b_k,
\qquad \lambda\in I=[\lambda_-,\lambda_+]
\label{eq:affine}
\end{equation}
Define $Q_k(\lambda_1,\lambda_2):=\mathbf d(k,\lambda_1)\cdot\mathbf d(k,\lambda_2)$. This function is affine in each argument, so its minimum on $I\times I$ lies at a corner. Since $Q_k(\lambda,\lambda)=|\mathbf d(k,\lambda)|^2>0$ inside the gapped interval, only the mixed corner is nontrivial. Therefore, if
\begin{equation}
\bigl(\mathbf a_k+\lambda_-\mathbf b_k\bigr)\cdot
\bigl(\mathbf a_k+\lambda_+\mathbf b_k\bigr)>0
\qquad \forall\,k,
\label{eq:affinecriterion}
\end{equation}
then every quench with endpoints in $I$ satisfies \eqref{eq:criterion}. The corollary gives a simple sufficient test for a whole interval of quenches, while the theorem gives the exact pairwise region in the full two-parameter plane.

\begin{figure}[H]
	\centering
	\includegraphics[width=0.9\linewidth]{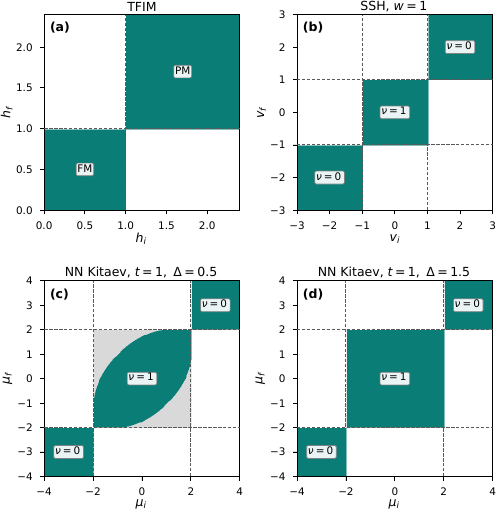}
	\caption{
		Physical same-phase regions and exact theorem-valid regions for four one-parameter free-fermion models. White denotes parameter pairs whose endpoints are not in the same connected gapped phase. Gray denotes same-phase quenches for which the ordering theorem fails. Teal denotes the exact theorem-valid region, where the final ground state is the unique maximal-overlap final eigenstate. Dashed lines mark equilibrium phase boundaries. Labels inside the diagonal blocks indicate the phase character: FM and PM for the TFIM, and winding number $\nu$ for the SSH and Kitaev chains. (a) TFIM for $h \ge 0$. The theorem-valid region coincides with the ferromagnetic and paramagnetic phases. (b) SSH chain with fixed $w=1$. The theorem-valid region coincides with the full physical phase diagram. (c) Nearest-neighbor Kitaev chain with $t=1$ and weak pairing $\Delta=0.5$. Inside the topological sector $\nu=1$, the theorem-valid region is a strict subset of the physical same-phase region. (d) Nearest-neighbor Kitaev chain with $t=1$ and strong pairing $\Delta=1.5$. The theorem-valid region again coincides with the full physical same-phase region.
	}
	\label{fig:compare}
\end{figure}

\paragraph*{TFIM chain.}
For the transverse-field Ising model after Jordan--Wigner and Bogoliubov transformation, we have
\begin{equation}
	\mathbf d(k;h)=\bigl(0,\sin k,h-\cos k\bigr)
	\label{eq:TFIM_d}
\end{equation}
which is affine in the transverse field $h$: $\mathbf d(k;h)=\mathbf a_k+h\,\mathbf b_k,
\; \mathbf a_k=(0,\sin k,-\cos k),\;
\mathbf b_k=(0,0,1).$
Then $	Q_k(h_-,h_+)=1+h_-h_+-(h_-+h_+)\cos k.$
For the physical convention $h\ge 0$, the minimum over the Brillouin zone occurs at $\cos k=1$, so $Q_{\min}=(1-h_-)(1-h_+).$
Hence the maximal connected theorem-valid intervals are
\begin{equation}
		\mathcal P_{\mathrm{FM}}=(0,1),
		\qquad
		\mathcal P_{\mathrm{PM}}=(1,\infty)
	\label{eq:TFIM_P}
\end{equation}
These are exactly the ferromagnetic (\amend{FM}) and paramagnetic (\amend{PM}) phases used by Damerow and Kehrein. This is also visually clear from the Fig. \ref{fig:compare}a

\paragraph*{SSH chain.}
For the SSH model, $ h(k)=\bigl(v+w\cos k\bigr)\sigma_x+w\sin k\,\sigma_y,$
where $v$ and $w$ are the intra-cell and inter-cell hoppings \cite{SSH1979}. For a quench $v_i\to v_f$ at fixed $w$: $Q_k(v_i,v_f)=\mathbf d_i\cdot\mathbf d_f=v_i v_f+w(v_i+v_f)\cos k+w^2$
Because $Q_k$ is linear in $\cos k$, its minimum occurs at $k=0$ or $k=\pi$:
\begin{equation}
\min_k Q_k(v_i,v_f)=
\min\!\bigl\{(v_i+w)(v_f+w),(v_i-w)(v_f-w)\bigr\}
\label{eq:sshmin}
\end{equation}
Hence, for fixed $w>0$, the exact theorem-valid region is the union of the three connected gapped sectors separated by the gap closings at $v=\pm w$. In particular, quenches with $|v_i|, |v_f|<w$ or $|v_i|, |v_f|>w$ have the final ground state as the unique maximal-overlap final eigenstate. Therefore the maximal connected theorem-valid intervals are
\begin{equation}
		\mathcal P_{\mathrm{top}}=(-w,w),
		\quad
		\mathcal P_{\mathrm{triv}}=(-\infty,-w)\cup(w,\infty)
	\label{eq:SSH_P}
\end{equation}
These coincide with the standard topological and trivial SSH phases on the positive-coupling line, also shown in Fig. \ref{fig:compare}b.

\paragraph*{Nearest-neighbor Kitaev chain.}
For the Kitaev chain with nearest-neighbor (\amend{NN}) hopping $t$ and pairing $\Delta$: $\mathbf d(k;\mu)=\bigl(0,2\Delta\sin k,-\mu-2t\cos k\bigr)$
with gap closings at $\mu=\pm 2t$ for $\Delta\neq0$ \cite{Kitaev2001}. For a chemical-potential quench,
\begin{align}
Q_k(\mu_i,\mu_f)
&=(\mu_i+2t\cos k)(\mu_f+2t\cos k)+4\Delta^2\sin^2 k \nonumber\\
&=\mu_i\mu_f+2t(\mu_i+\mu_f)\cos k+4\Delta^2\nonumber\\
&\hspace{2.5cm}+4(t^2-\Delta^2)\cos^2 k.
\label{eq:NNKitaevQ}
\end{align}
This admits a complete analytic classification.

For the two outer gapped intervals, $\mu_i,\mu_f>2t\;\text{or} \;\mu_i,\mu_f<-2t,$
the factors $\mu_\alpha+2t\cos k$ keep fixed sign for all $k\in[0,\pi]$. Hence
\begin{equation}
	Q_k(\mu_i,\mu_f)>0 \qquad \forall\,k
\end{equation}
for any $\Delta\neq0$. Therefore the theorem-valid region coincides with the physical same-phase region in the two trivial sectors (see, Fig. \ref{fig:compare}c) for arbitrary pairing strength.

The topological interval $-2t<\mu<2t$ depends on the ratio $|\Delta|/t$. If $|\Delta|\ge t$, then the coefficient of $\cos^2 k$ in \eqref{eq:NNKitaevQ} is nonpositive, so $Q_k$ is concave as a function of $x=\cos k\in[-1,1]$. The minimum is therefore reached at $x=\pm1$, and the exact criterion reduces to $(\mu_i\pm 2t)(\mu_f\pm 2t)>0.$
Thus the full topological interval is theorem valid in the strong-pairing regime.

For $0<|\Delta|<t$, by contrast, $Q_k$ is convex in $x=\cos k$. The stationary point lies at
\begin{equation}
	x_\ast=-\frac{t(\mu_i+\mu_f)}{4(t^2-\Delta^2)}
	\label{eq:xstar}
\end{equation}
If $|x_\ast|\geq1$, the minimum is again attained at $x=\pm1$, so the same-phase condition remains sufficient. The only nontrivial case is $|x_\ast|\le1$, that is,
\begin{equation}
	|\mu_i+\mu_f|\le \frac{4(t^2-\Delta^2)}{t}
	\label{eq:strip}
\end{equation}
Then the minimum occurs at $x_\ast$. Hence, inside the strip \eqref{eq:strip}, the theorem-valid region is the exact ellipse
\begin{equation}
	\frac{(\mu_i+\mu_f)^2}{16(t^2-\Delta^2)}
	+
	\frac{(\mu_i-\mu_f)^2}{16\Delta^2}
	<1
	\label{eq:NNellipse}
\end{equation}
Therefore same-phase violations already occur in the nearest-neighbor Kitaev chain when $0<|\Delta|<t$. A simple example is the symmetric quench: $	\mu_i=-m,\; \mu_f=m,\; 2|\Delta|<|m|<2t,$
for which $k=\pi/2$ gives
\begin{equation}
	Q_{\pi/2}=4\Delta^2-m^2<0
\end{equation}
The final ground state is then not the maximal-overlap final eigenstate although both endpoints lie in the same connected topological phase. Fig. \ref{fig:compare}c,d shows the exact theorem-valid region for the weak-pairing example $t=1$, $\Delta=0.5$. In the strong-pairing line $t=1 \text{ and }\Delta=1.5$, the full same-phase region is recovered.

\paragraph*{Finite-range Kitaev chain.} Above, we have seen the violation of the original conjecture \cite{DamerowKehrein2025} in the weak pairing limit in NN Kitaev chain. Now we investigate the violation in the finite-range spinless superconductor
\begin{multline}
H_R(\mu)=\sum_{r=1}^{R}\sum_j
\Bigl[
-t_r c_j^\dagger c_{j+r}+\Delta_r c_j c_{j+r}+\hc
\Bigr]\\
-\mu\sum_j\left(c_j^\dagger c_j-\frac12\right)
\label{eq:HR}
\end{multline}
with real $t_r$ and $\Delta_r$. After Fourier transform and Nambu doubling, $\Psi_k=(c_k,c_{-k}^\dagger)^T$, the Hamiltonian becomes
\begin{equation}
H_R(\mu)=\sum_{0<k<\pi}\Psi_k^\dagger
\bigl[d_y(k)\tau_y+d_z(k;\mu)\tau_z\bigr]\Psi_k
\label{eq:bdgR}
\end{equation}
where, $d_y(k)=2P_R(k)$, and $d_z(k;\mu)=-\mu-2T_R(k)$ 
with $T_R(k)=\sum_{r=1}^{R}t_r\cos(rk),
\quad
P_R(k)=\sum_{r=1}^{R}\Delta_r\sin(rk).$
Thus the Bloch vector is $\mathbf d_R(k;\mu)=\bigl(0,2P_R(k),-\mu-2T_R(k)\bigr),$
and the exact criterion becomes
\begin{equation}
Q_R(k;\mu_i,\mu_f):=
(\mu_i+2T_R(k))(\mu_f+2T_R(k))+4P_R(k)^2>0
\label{eq:QR}
\end{equation}
for all $k$. The physical phase boundaries are given by the bulk-gap condition $P_R(k)=0$ together with $\mu+2T_R(k)=0$.
Now choose the illustrative family $R=3$ with $t_r=\Delta_r=1$. Then
$T_3(k)=\cos k+\cos2k+\cos3k,\; P_3(k)=\sin k+\sin2k+\sin3k$
Using trigonometric identities,
\begin{equation}
P_3(k)=\sin 2k\,(1+2\cos k)
\label{eq:P3factor}
\end{equation}
so $P_3(k)=0$ at $k=0, \pm\pi/2,\pm2\pi/3$. Evaluating $T_3(k)$ at these points gives: $T_3(0)=3,\; T_3(\pm\pi/2)=-1,\; T_3(\pm 2\pi/3)=0.$
Hence the bulk gap closes at $\mu=-6, 2, 0.$
The physical line therefore splits into four connected gapped regions, $(-\infty,-6),\; (-6,0),\; (0,2),\text{and } (2,\infty),$
with winding numbers $\nu=0,-1,-3,0$, respectively.

\begin{figure}[!htp]
	\centering
	\includegraphics[width=0.77\columnwidth]{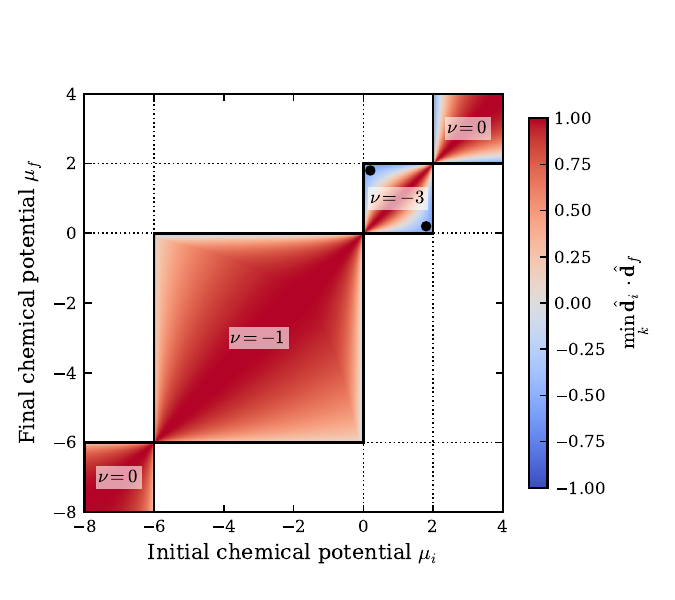}
	\caption{Detailed validity map for the finite-range Kitaev chain with $R=3$ and $t_r=\Delta_r=1$ for $1\le r\le 3$. The color scale shows $\min_k\,\dk(k;\mu_i)\cdot\dk(k;\mu_f)$. Red means that the theorem holds, blue means that it fails, and the white contour gives the exact boundary. White regions are excluded because the two endpoints are not in the same connected gapped region. Dashed lines mark the bulk gap closings at $\mu=-6,0,2$.}
	\label{fig:R3map}
\end{figure}

The exact theorem-valid region is obtained from \eqref{eq:QR} by requiring $\min_k\,\dk(k;\mu_i)\cdot\dk(k;\mu_f)>0.$
Fig. \ref{fig:R3map} shows the full same-phase domain together with the exact sign of this minimum. The theorem-valid region is strictly smaller than the physical same-phase region. The difference is not a small boundary effect. It extends over a finite area inside the phase box $0<\mu<2$, and it also appears in the phase $2<\mu<6$.

A concrete same-phase counterexample is $\mu_i=0.2,\quad \mu_f=1.8,$
which lies fully inside the phase $0<\mu<2$. For this quench, $\min_k\,\dk(k;0.2)\cdot\dk(k;1.8)\approx -0.478992$
at $k\approx1.75349$. Therefore the final ground state is not the maximal-overlap final eigenstate. This is an explicit same-phase violation of the phase-only rule inside an exactly solvable free-fermion family.

\paragraph*{DQPTs consequences.}
The exact overlap-ordering theorem has an immediate dynamical consequence in every factorized free-fermion system. For a quench from the initial ground state $\ket{\Psi_{0,i}}$ to a final Hamiltonian $H_f$, the Loschmidt amplitude is $\mathcal G(t)=\bra{\Psi_{0,i}}e^{-iH_f t}\ket{\Psi_{0,i}}$, and the standard DQPT diagnostic\cite{HeylPolkovnikovKehrein2013} is the return-rate density $r(t)=-(1/L)\ln |\mathcal G(t)|^2$. In a factorized two-level description, $\mathcal G(t)=\prod_k g_k(t)$, where each sector is controlled by the final quasiparticle energy $E_k^f$ and the sectorwise Bloch-vector overlap $x_k=\hat{\mathbf d}_{i,k}\!\cdot\!\hat{\mathbf d}_{f,k}$. Since the initial sector ground state has final-basis weights $p_k^-=(1+x_k)/2$ and $p_k^+=(1-x_k)/2$, the sector Loschmidt amplitude is

\begin{equation}
	g_k(t)=\frac{1+x_k}{2}e^{iE_k^f t}+\frac{1-x_k}{2}e^{-iE_k^f t}\label{eq:gk}
\end{equation}

This gives $|g_k(t)|^2=\cos^2(E_k^f t)+x_k^2\sin^2(E_k^f t)$, so a real-time zero can occur only if $x_k=0$, in which case the critical times are $t_{n,k}^*=(2n+1)\pi/(2E_k^f)$. Equivalently, the Fisher zeros $z_{n,k}$ of the analytic continuation $\mathcal G(z)$ are $z_{n,k}=[\ln((1-x_k)/(1+x_k))+i\pi(2n+1)]/(2E_k^f)$, and they cross the real-time axis precisely when $x_k=0$. Therefore the theorem-valid condition $x_k>0$ for all $k$ implies that no Fisher-zero line can hit the real axis and hence that no DQPT occurs. In this class, DQPTs are the exact dynamical signature of the breakdown of overlap ordering: $x_k>0$ for all $k$ implies ground-state dominance and no DQPT, while the onset of a DQPT is controlled by the first critical momentum $k^*$ for which $x_{k^*}=0$, or equivalently $p_{k^*}=1/2$.

\begin{figure}[H]
	\centering
	\includegraphics[width=\textwidth]{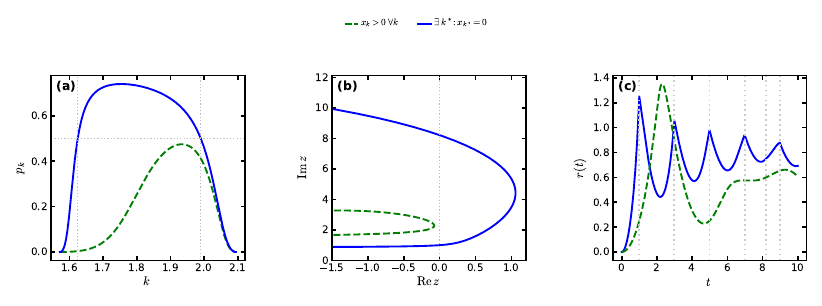}
	\caption{\label{fig:R3_dqpt_combined}
		DQPT diagnostics for two same-phase quenches of the finite-range $R=3$ Kitaev chain with $t_r=\Delta_r=1$ for $r=1,2,3$, both lying inside the connected equilibrium phase $0<\mu<2$ (i.e. $\pi/2<k<2\pi/3$). Dashed green curves correspond to the theorem-valid quench $(\mu_i,\mu_f)=(0.2,1.0)$, for which $x_k>0$ for all $k$. Solid blue curves correspond to the theorem-invalid same-phase quench $(\mu_i,\mu_f)=(0.2,1.8)$, for which $x_k=0$ at the two critical momenta $k^*\approx 1.62401$ and $k^*\approx 1.98970$. (a) Mode-resolved excitation probability $p_k=(1-x_k)/2$; the dotted horizontal line marks the threshold $p_k=1/2$, equivalent to $x_k=0$, and the vertical dotted lines mark the critical modes. (b) Lowest Fisher-zero line $z_{0,k}$; the vertical dotted line marks the real-time axis $\mathrm{Re}\,z=0$. (c) Return-rate density $r(t)$, with vertical dotted lines marking the critical times $t_n^*=(2n+1)\pi/(2E_{k^*}^f)$ obtained from the Fisher-zero crossings. The figure shows that, although both quenches remain in the same connected equilibrium phase, only the theorem-valid one is DQPT-free, while the theorem-invalid one exhibits the standard sequence of Fisher-zero crossings and cusp singularities.}
\end{figure}

This mechanism is especially transparent in the finite-range $R=3$ Kitaev chain \eqref{eq:HR}. The bulk gap closes at $\mu=-6,0,2$, so $0<\mu<2$ is one connected physical phase. However, the theorem-valid region inside this phase is only the subset for which $\min_k x_k>0$. Thus, the same physical phase splits into a theorem-valid subregion and a theorem-invalid subregion. In the former, $p_k=(1-x_k)/2<1/2$ for all $k$, the Fisher zeros remain off the real axis, and $r(t)$ stays smooth. In the latter, there exists at least one critical mode $k^*$ with $x_{k^*}=0$, so $p_{k^*}=1/2$ (see, Fig. \ref{fig:R3_dqpt_combined}a), the Fisher-zero line crosses the real axis at $t_n^*=(2n+1)\pi/(2E_{k^*}^f)$ (see, Fig. \ref{fig:R3_dqpt_combined}b), and $r(t)$ develops the nonanalytic cusps characteristic of a DQPT (see, Fig. \ref{fig:R3_dqpt_combined}c). This provides a same-phase DQPT mechanism that is absent in\cite{HeylPolkovnikovKehrein2013} and directly ties the dynamical transition to the exact overlap criterion.

\onecolumngrid

\twocolumngrid

\paragraph*{Discussion.}
The result is controlled by a sectorwise geometric criterion rather than by phase continuity. In factorized free-fermion systems, overlap ordering is determined by the sign of $\hat{\mathbf d}_{i,k}\!\cdot\!\hat{\mathbf d}_{f,k}$ in each sector. This explains the TFIM and SSH cases, where the theorem-valid region matches the full same-phase region, and also identifies where this correspondence fails. In the nearest-neighbor Kitaev chain it holds only for $|\Delta|\ge t$, whereas for $0<|\Delta|<t$ and in the finite-range Kitaev chain the same physical phase can contain theorem-invalid quenches. The same criterion also controls the dynamics: $x_k>0$ for all $k$ excludes Fisher-zero crossings and hence DQPTs, while leaving the theorem-valid region allows same-phase DQPTs through critical modes with $x_k=0$. \amend{The exact geometric criterion relies on sector factorization. Once, we introduce a generic interaction, we loose the analytical control, and therefore our theorem is not expected to hold in a generic interacting system. However, at weak coupling in a gapped regime, a mean-field description may provide an approximate diagnostic, provided the dressed dot products remain separated from zero by a finite margin. We could validate such diagnostics using numerical benchmark such as exact-diagonalization method.}

\paragraph*{Conclusion.}
We derived an exact necessary-and-sufficient criterion for ground-state overlap dominance after sudden quenches in translationally invariant factorized class of free-fermion systems. The final ground state is the unique maximal-overlap final eigenstate iff all sectorwise Bloch-vector overlaps are positive. This yields exact model comparisons with physical phase regions, explicit same-phase counterexamples to a phase-only rule in the Kitaev chain, and a direct dynamical corollary: theorem-valid quenches are DQPT-free, whereas theorem-invalid same-phase quenches can exhibit DQPTs. The main open question is which additional (if any) structures, beyond factorization, can make phase continuity sufficient.

\paragraph*{Code availability.}
All the codes that were used to generate figures in this work are available from the author upon reasonable request.

\bibliographystyle{apsrev4-2}

%

\end{document}